\documentclass[aps,pra,showpacs]{revtex4}

\usepackage{graphicx, amsmath, amsfonts, amssymb}

\begin{document}

\title{Quasi-exact solution to the Dirac equation for the hyperbolic secant potential}

\author{R. R. Hartmann}
\email[]{richard.hartmann@dlsu.edu.ph}
\affiliation{
Physics Department,
De La Salle University
2401 Taft Avenue,
Manila,
Philippines
}

\author{M. E. Portnoi}
\email[]{m.e.portnoi@exeter.ac.uk}
\affiliation{School of Physics, University of Exeter, Stocker Road, Exeter EX4 4QL, United Kingdom\\
and International Institute of Physics, Av. Odilon Gomes de Lima, 1722, Capim Macio, CEP: 59078-400, Natal - RN, Brazil.
}

\date{6 November 2013}

\begin{abstract}
We analyze bound modes of two-dimensional
massless Dirac fermions confined within a hyperbolic secant potential, which
provides a good fit for potential profiles of existing top-gated graphene
structures. We show that bound states of both positive and negative
energies exist in the energy spectrum and that there is a threshold
value of the characteristic potential strength for which the first
mode appears. Analytical solutions are presented in several limited cases and supercriticality is discussed.

\end{abstract}



\maketitle

\section{Introduction}
Transmission resonances and supercriticality \cite{Dombey_PRL_20} (bound states occurring at $E=-m$, where $E$ is the particles energy and $m$ the particles mass)
of relativistic particles in one-dimensional potential wells have been studied extensively \cite{Dombey_PRL_20,Coulter_AJP_71,Calogeracos_PN_96,Kennedy_JPA_02,Villalba_PRA_03,Guo_EJP_09,Sogut_PS_11,Arda_PS_11,Villalba_PS_10,Kennedy_IJMPA_04}. Analytic solutions have been obtained for the the square well \cite{Coulter_AJP_71,Calogeracos_PN_96},
Woods-Saxon potential \cite{Kennedy_JPA_02}, cusp potential \cite{Villalba_PRA_03}, Hulth\'{e}n potential \cite{Guo_EJP_09} as well as asymmetric barriers \cite{Sogut_PS_11,Arda_PS_11}, multiple barriers \cite{Villalba_PS_10} and a class of short-range potentials \cite{Kennedy_IJMPA_04}. The successful isolation of graphene \cite{Novoselov_04} has led to renewed interest in the transmission-reflection problem for the one-dimensional Dirac equation.

The carriers within graphene, a single layer of carbon atoms in a honeycomb lattice, behave as two-dimensional massless Dirac fermions \cite{CastroNeto_Rmp_09}. In the presence of an electric field, their massless relativistic nature results in drastically different behavior to their normal non-relativistic electron counterparts, for example, backscattering is forbidden for carriers which are incident normal to the barrier \cite{Klein,Ando_98,Katsnelson_NP_06}. However, they can be reflected at non-normal incidence and therefore confinement is possible.

Electron waveguides in graphene have been studied extensively both theoretically and experimentally. It has been shown that it is possible to confine graphene electrons by electrostatic potentials \cite{Chaplik_06,Peeters_PRB_06,Chau_PRB_09,Zhang_APL_09,Titov_PRL_09,Williams_NanoT_11,Yuan_JAP_11,Wu_APL_11,Downing_PRB_11,Ping_12_CTP,Katsnelson_PS_12,Miserev_JETP_12,Katsnelson_AoP_13}, magnetic barriers
\cite{Pereira_PRB_07,Martino_PRL_07,Martino_SSC_07,Shytov_PR_08,DellAnna_PRB_09,Ghosh_JPCM_09,Ghosh_JPCM_09_2,Kuru_JPCM_09,Sharma_JPCM_11,Myoung_PRB_11,Huang_JAP_12}
 and strain-induced fields \cite{Low_NL_08,Pereira_PRL_09,Guinea_NATP_10,Wu_PTL_11}. Transmission through symmetric \cite{Katsnelson_NP_06,Cheianov_PRB_06,Peeters_PRB_06,Chaplik_06,PeetersAPL07,Zhang_PRL_08,Shytov_PR_08, Fogler_PRB_08,BeenakkerRMP08,Chau_PRB_09,TransportPN,Klein+TopGate,TopGate1+Klein,Klein+TopGate1,riverside,Goldhaber-Gordon,Klein+Topgate2,Miserev_JETP_12,Katsnelson_AoP_13,Mouhafid_13} and asymmetric electrostatic barriers \cite{Ping_12_CTP} have been studied and fully confined modes within a smooth one-dimensional potential have been predicted to exist at zero-energy \cite{Hartmann_1,Downing_1}. The majority of electrostatically defined waveguides have been limited to sharp barriers (i.e. potentials which are non-continuous). However, unlike in semiconductor heterostructures or dielectric waveguides for light, finite square wells and other sharply-terminated finite barriers have not yet been experimentally demonstrated in graphene, as potential profiles are created by electrostatic gating which results in smooth potentials \cite{TransportPN,Klein+TopGate,TopGate1+Klein,Klein+TopGate1,riverside,Goldhaber-Gordon,Klein+Topgate2}.


The conduction and valence bands in graphene touch each other at six points, which lie on the edge of the first Brillouin zone. In pristine undoped graphene the Fermi surface coincides with these points (known as Dirac points) and at these points the dispersion relation is linear \cite{Wallace_Phys_Rev_47}. Two of these points are inequivalent and degenerate in terms of energy.
A sharp barrier results in intervalley scattering, therefore the full treatment of a sharp boundary requires the mixing of two Dirac cones. Therefore to stay within a single cone approximation many authors introduce the term ``smooth step-like potential", since smooth potentials changing slowly on the spatial scale exceeding tens of graphene lattice constants, like the one considered in this paper, do not result in intervalley scattering. Furthermore, for sharp barriers, the discontinuity of the potential results in discontinuities in the wavefunction's derivative which have to be treated with special care \cite{Dragoman_09}, this is not the case for smooth potentials.


Exact solutions of the one-dimensional Dirac equation are not only useful in the analytic modeling of physical systems, but they are also important for testing numerical, perturbative or semi-classical methods \cite{Katsnelson_AoP_13}. The hyperbolic secant potential belongs to the class of quantum models which are quasi-exactly solvable \cite{Turbiner_JETP_88,Ushveridze_94,Bender_JPA_98,Downing_JMP_13}, where only some of the eigenfunctions and eigenvalues are found explicitly.

In this paper we obtain the bound state energies contained within the hyperbolic secant potential in pristine
graphene and supercriticality is discussed. Hitherto unknown analytical solutions for certain bound modes contained within this model potential are presented. We show that bound states of both positive and negative energies exist in the spectrum and that there is a threshold value of the characteristic potential strength for which the first mode appears, in striking contrast to the non-relativistic case.

\section{Bound modes in a model potential}

The Hamiltonian operator in the massless Dirac-Weyl model for graphene,
which describes the motion of a single electron in the presence of
a one-dimensional potential $U\left(x\right)$ is
\begin{equation}
\hat{H}=v_{\mathrm{F}}\left(\sigma_{x}\hat{p}_{x}+\sigma_{y}\hat{p}_{y}\right)+U\left(x\right),
\label{eq:Ham_orig}
\end{equation}
where $\sigma_{x,y}$ are the Pauli spin matrices, $\hat{p}_{x}=-i\hbar\frac{\partial}{\partial x}$
and $\hat{p}_{y}=-i\hbar\frac{\partial}{\partial y}$ are the momentum
operators in the $x$ and $y$ directions respectively and $v_{\mathrm{F}}\approx 10^{6}$~m/s
is the Fermi velocity in graphene. In what follows we will consider a smooth confining potential, the hyperbolic secant potential, which does not mix the two non-equivalent valleys. All our results herein can be easily reproduced for the other
valley. When Eq.~(\ref{eq:Ham_orig}) is applied to a two-component
Dirac wavefunction of the form:
\[
e^{ik_{y}y}\left(\begin{array}{c}
\Psi_{A}\left(x\right)\\
\Psi_{B}\left(x\right)
\end{array}\right),
\]
where $\Psi_{A}\left(x\right)$ and $\Psi_{B}\left(x\right)$ are
the wavefunctions associated with the $A$ and $B$ sublattices of
graphene respectively and the free motion in the $y$-direction is
characterized by the wave vector $k_{y}$ measured with respect to
the Dirac point, the following coupled first-order differential equations
are obtained:
\begin{equation}
\left(V\left(x\right)-\varepsilon\right)\Psi_{A}-i\left(\frac{d}{dx}+k_{y}\right)\Psi_{B}=0
\label{eq:ham_1}
\end{equation}
and
\begin{equation}
\left(V\left(x\right)-\varepsilon\right)\Psi_{B}-i\left(\frac{d}{dx}-k_{y}\right)\Psi_{A}=0.
\label{eq:ham_2}
\end{equation}
Here $V\left(x\right)=U\left(x\right)/\hbar v_{\mathrm{F}}$ and energy,
$\varepsilon$, is measured in units of $\hbar v_{\mathrm{F}}$. For
convenience let $\Psi_{A}=\left(\Psi_{1}+\Psi_{2}\right)/2$ and $\Psi_{B}=\left(\Psi_{1}-\Psi_{2}\right)/2$
therefore Eqs.~(\ref{eq:ham_1}-\ref{eq:ham_2}) become
\begin{equation}
\left(V\left(x\right)-\varepsilon-i\frac{d}{dx}\right)\Psi_{1}+ik_{y}\Psi_{2}=0
\label{eq:ham_3}
\end{equation}
and
\begin{equation}
\left(V\left(x\right)-\varepsilon+i\frac{d}{dx}\right)\Psi_{2}-ik_{y}\Psi_{1}=0.
\label{eq:ham_4}
\end{equation}
Eqs.~(\ref{eq:ham_3}-\ref{eq:ham_4}) can then be reduced to a single
second-order differential equation in $\Psi_{1}$ $\left(\Psi_{2}\right)$
\begin{equation}
\left[\left(V\left(x\right)-\varepsilon\right)^{2}-k_{y}^{2}\pm i\frac{dV\left(x\right)}{dx}\right]\Psi_{1,2}+\frac{d^{2}\Psi_{1,2}}{dx^{2}}=0.
\label{eq:second_order_org}
\end{equation}
The plus and minus signs corresponds to wavefunction $\Psi_{1}$ and
$\Psi_{2}$ respectively. The potential under consideration is defined as
\begin{equation}
V\left(x\right)=-\frac{V_{0}}{\cosh\left(x/l\right)},
\label{eq:potential}
\end{equation}
where $V_{0}$ and $l$ characterize the potential strength and width
respectively. This potential is known to admit analytic solutions
for the case of $\varepsilon=0$ \cite{Hartmann_1,Hartmann_2} and is a good representation
of experimentally generated potential profiles \cite{Klein+TopGate,TopGate1+Klein,Klein+TopGate1,riverside,Goldhaber-Gordon,Klein+Topgate2}.
For top gated structures, the width of the potential is defined by the geometry of
the top gate structure, and the strength of the potential is defined
by the voltage applied to the top gate.

It should be noted that many unusual situations may arise in one-dimensional quantum mechanics due to the presence of a delta function when the usual definition does not hold true. In certain instances, the one-dimensional Dirac equation, which has a wavefunction defined by a differential equation involving the delta function, can result in the usual definition of the delta function being inconsistent with the definition of the wavefunction itself \cite{Sutherland_PRA81,Calkin_AJP_87,McKellar_PRC_87}. The implications of such situations regarding the transmission-reflection problem in one-dimensional quantum mechanics are reviewed at length in \cite{Coutinho_09}. In the limit that $l\rightarrow0$ the hyperbolic secant potential smoothly approaches a delta-function potential, thus making it an ideal approximation.

Let us search for solutions of Eq.~(\ref{eq:second_order_org}) with the potential given by Eq.~(\ref{eq:potential}) in the form
\begin{equation}
\Psi_{1,2}=A_{1,2}V^{\kappa}\psi_{1,2}\left(x\right),
\label{eq:first_wave}
\end{equation}
where
\begin{equation}
\kappa=l\sqrt{k_{y}^{2}-\varepsilon^{2}}
\label{eq:kappa}
\end{equation}
and $A_{1,2}$ is a constant. Substitution of Eq.~(\ref{eq:first_wave}) into Eq.~(\ref{eq:second_order_org})
yields
\[
\frac{d^{2}\psi_{1,2}}{dz^{2}}-8\kappa\frac{\tanh\left(z\right)}{1+\tanh^{2}\left(z\right)}\frac{d\psi_{1,2}}{dz}
\]
\begin{equation}
+4\left[2w\left(\mathrm{S}_{E}\sqrt{\Delta^{2}-\kappa^{2}}+\mathrm{S}_{\Psi}\frac{\tanh\left(z\right)}
{1+\tanh^{2}\left(z\right)}i\right)\frac{1-\tanh^{2}\left(z\right)}{1+\tanh^{2}\left(z\right)}+
\left[w^{2}-\kappa\left(\kappa+1\right)\right]\left(\frac{1-\tanh^{2}\left(z\right)}
{1+\tanh^{2}\left(z\right)}\right)^{2}\right]\psi_{1,2}=0,
\label{eq:diff_y}
\end{equation}
where we use the dimensionless variables $w=V_{0}l$, $\Delta=k_{y}l$,
$E=\varepsilon l$ and $z=x/2l$. $\mathrm{S}_{E}=1$ for $E>0$ and
$\mathrm{S}_{E}=-1$ for $E<0$. $\mathrm{S}_{\Psi}=1$ for $\Psi_{1}$
and $\mathrm{S}_{\Psi}=-1$ for $\Psi_{2}$. Using the transformation
$\psi_{1,2}=\left(\xi-\frac{1}{2}\right)^{\mu}H_{1,2}\left(\xi\right)$
with the change of variable
\[
\xi=\frac{e^{-i\frac{\pi}{4}}}{\sqrt{2}}
\frac{\tanh\left(z\right)+1}{\tanh\left(z\right)-i},
\]
where
\begin{equation}
\mu=\kappa+\mathrm{S}_{\mu}w+\frac{1}{2}\left(1+\mathrm{S}_{\mu}\mathrm{S}_{\Psi}\right)\label{eq:mew}
\end{equation}
and $\mathrm{S}_{\mu}=\pm1$, allows Eq.~(\ref{eq:diff_y}) to be
reduced to
\begin{equation}
\frac{d^{2}H_{1,2}}{d\xi^{2}}+\left[\frac{\gamma}{\xi}
+\frac{\delta}{\xi-1}+\frac{\epsilon}{\xi-a}\right]\frac{dH_{1,2}}{d\xi}
+\frac{\alpha\beta\xi-q}{\xi\left(\xi-1\right)\left(\xi-a\right)}H_{1,2}=0,
\end{equation}
where
\begin{eqnarray}
\nonumber
\epsilon &=& \alpha+\beta-\gamma-\delta+1,\\
\nonumber
\gamma &=& \delta=1+2\kappa,\\
\nonumber
a &=& \frac{1}{2},\\
\nonumber
\beta &=& \frac{2}{\alpha}\left[S_{\Psi}w+\left(1+2\kappa\right)\mu\right],\\
\nonumber
2\alpha &=& 2\mu+2\kappa+1\pm\sqrt{\left(2\mu+2\kappa+1\right)^{2}-8\left[\mathrm{S}_{\Psi}w+\left(1+2\kappa\right)\mu\right]},\\
\nonumber
q&=&\frac{\alpha\beta}{2}+i2\mathrm{S}_{E}w\sqrt{\Delta^{2}-\kappa^{2}},\\
\nonumber
\end{eqnarray}
and $H_{1,2}$ is the Heun function given by the expression \cite{Heun_89}
\begin{equation}
H_{1,2}=H_{1,2}\left(a,\, q;\,\alpha,\,\beta,\,\gamma,\,\delta,\,\xi\right)=\sum_{j=0}^{\infty}c_{j}\xi^{j},
\label{eq:Heun_fun}
\end{equation}
where
\begin{equation}
c_{0}=1,\qquad a\gamma c_{1}-qc_{0}=0,\qquad R_{j}c_{j+1}-\left(Q_{j}+q\right)c_{j}+P_{j}c_{j-1}=0,
\nonumber
\end{equation}
with
\begin{eqnarray}
\nonumber
P_{j}&=&\left(j-1+\alpha\right)\left(j-1+\beta\right),\\
\nonumber
Q_{j}&=&j\left[\left(j-1+\gamma\right)\left(1+a\right)+a\delta+\epsilon\right],\\
\nonumber
R_{j}&=&a\left(j+1\right)\left(j+\gamma\right).
\nonumber
\end{eqnarray}
The bound-state energies are determined by the boundary condition $\Psi_{1,2}\left(\pm\infty\right)=0$. For the case of $x\rightarrow-\infty$, $\xi\rightarrow0$ and the right hand side of Eq.~(\ref{eq:Heun_fun}) equals unity; therefore, it can be seen from Eq.~(\ref{eq:first_wave}) that the boundary condition $\Psi_{1,2}\left(-\infty\right)=0$ requires $\kappa$ to exceed zero. In the limit that $x\rightarrow\infty$, $\xi\rightarrow1$ the bound state solutions correspond to combinations of the accessory and exponent parameters which result in non-divergent values of Eq.~(\ref{eq:Heun_fun}). In certain instances Eq. (\ref{eq:Heun_fun}) can be reduced to a finite polynomial of degree $n$ admitting analytic results. However, applying symmetry conditions to the wavefunction Eq.~(\ref{eq:first_wave}) is sufficient to obtain the energy eigenvalue spectrum.

\begin{figure}[h]
\includegraphics*[height=100mm]{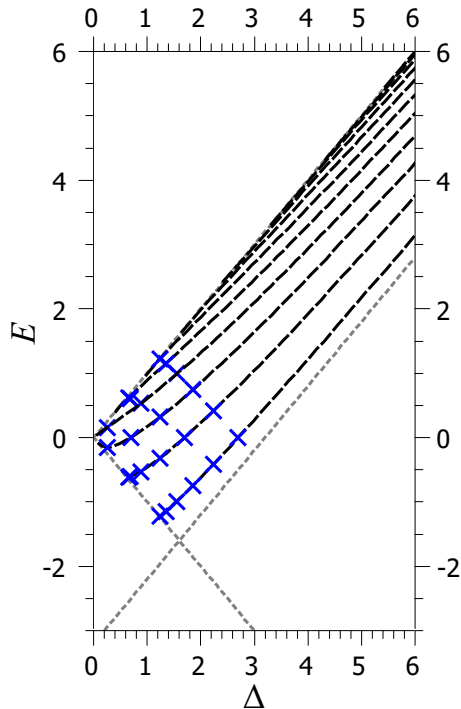}
\caption{ (Color online)
Energy spectrum of confined states in a hyperbolic-secant potential as a function of $\Delta$ for $w=3.2$. The black (long-dashed) lines represent the computational results and the blue crosses the analytical results. The boundary at which the bound states merge with the continuum is denoted by the grey (short-dashed) lines.
}
\label{fig_dispersion}
\end{figure}

\begin{figure}[h]
\includegraphics*[height=100mm,angle=270]{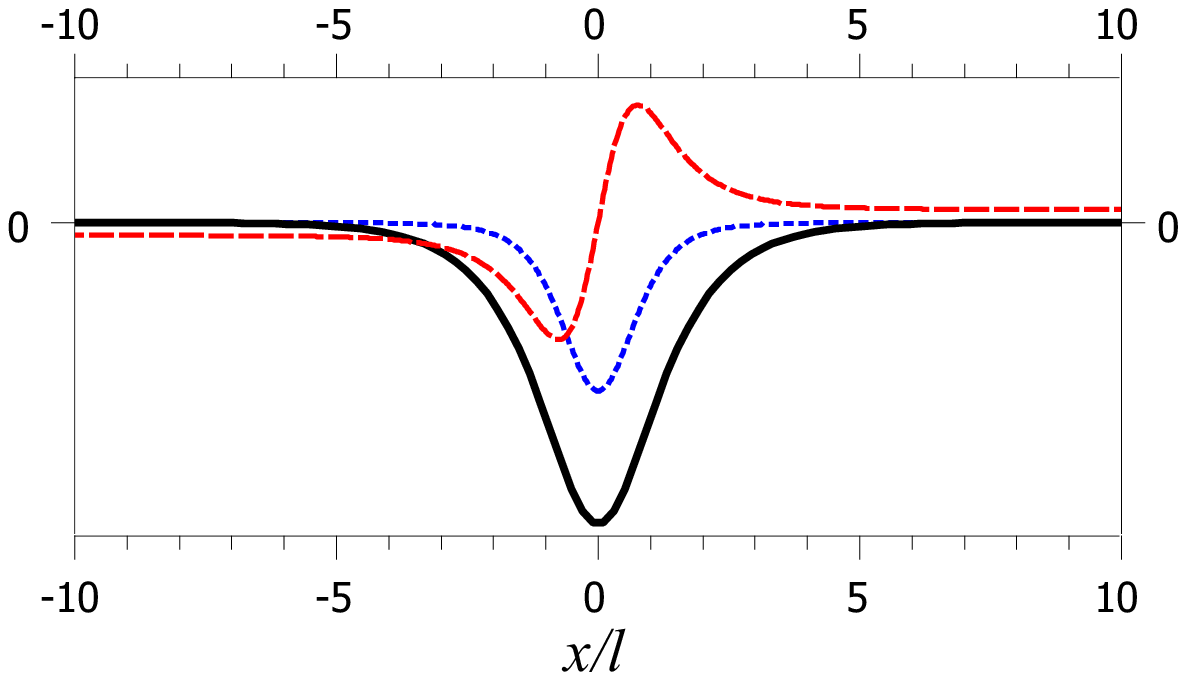}
\caption{(Color online)
The lowest energy, zero transverse momentum wave function for the hyperbolic potential well (depicted by the black solid line) defined by $\omega=3.2$. The real part of the wave functions $\Psi_{\mathrm{I}}$ and $\Psi_{\mathrm{II}}$ are shown in red (long-dashed) and blue (short-dashed) respectively.
}
\label{fig_super}
\end{figure}

It is clear from Eqs.~(\ref{eq:ham_3}-\ref{eq:ham_4}) that neither $\Psi_{1}$ nor $\Psi_{2}$ are symmetrized wavefunctions, so we shall transform to the symmetrized functions:
\begin{equation}
\Psi_{\mathrm{I}}=\left(\Psi_{1}+\Psi_{2}\right)+i\left(\Psi_{1}-\Psi_{2}\right),\qquad\Psi_{\mathrm{II}}=\left(\Psi_{1}+\Psi_{2}\right)-i\left(\Psi_{1}-\Psi_{2}\right).
\end{equation}
When $\Psi_{\mathrm{I}}$ is an odd function $\Psi_{\mathrm{I}}\left(0\right)=0$ and when $\Psi_{\mathrm{I}}$ is an even function $\Psi_{\mathrm{II}}\left(0\right)=0$ these two boundary conditions result in the following transendental equation:
\begin{equation}
2\left[\left(1+\mathrm{S}_{\Psi}\mathrm{S}_{\mu}\right)w
+\frac{1}{2}\left(\mathrm{S}_{\Psi}+\mathrm{S}_{\mu}\right)
+E\mp\Delta\right]H_{1,2}\left(\xi_{0}\right)
+i\mathrm{S}_{\Psi}\left.\frac{dH_{1,2}\left(\xi\right)}{d\xi}\right|_{\xi=\xi_{0}}=0,
\label{eq:transedental}
\end{equation}
where $\xi_{0}=e^{i\frac{\pi}{4}}/\sqrt{2}$ and the $\mp$ sign corresponds to the odd (upper sign) and even (lower sign) bound modes. Eq.~(\ref{eq:transedental}) was solved numerically and the results are shown in Fig.~\ref{fig_dispersion} for the case of $w=3.2$. The long-dashed lines represent $\kappa=0$; as $\kappa\rightarrow0$, $\left|E\right|\rightarrow\left|\Delta\right|$ and the bound states merge with the continuum, and the potential is said to be supercritical, where $\Delta$ plays the role of mass.  The zero transverse momentum wavefunctions (i.e. $\kappa=0$) are half-bound; one component of the spinor wavefunction decays to zero at $x=\pm\infty$ and the other component decays to non-zero values. The half-bound states can be obtained from Eq.~(\ref{eq:transedental}) with the substitution $E=\pm \Delta$. For the case of $w=3.2$, the first half-bound state occurs at $E=-1.232$ and the corresponding wave function is plotted in Fig.~\ref{fig_super}. It should be noted that this wavefunction contains two additional stationary points in comparison to that of the square well \cite{Dombey_PRL_20}; however, they are indeed present for a Gaussian \cite{Dombey_PRL_20} and Woods-Saxon potential \cite{Kennedy_JPA_02}.

\subsection{Exact solutions}
In what follows we shall see that Eq.~(\ref{eq:transedental}) can be solved exactly in a few limited cases. One condition that ensures $\Psi_{1}$ is a non-divergent function at $x\rightarrow\infty$ is that $H_{1,2}$ is reduced to a finite polynomial. This
occurs when two conditions are met:
\begin{equation}
\alpha=-n
\label{eq:cond_1}
\end{equation}
and
\begin{equation}
q=q_{n,m},
\label{eq:cond_2}
\end{equation}
where $n$ and $m$ are non-negative integers and $m\leq n$, and
$q_{n,m}$ are the eigenvalues of the tridiagonal matrix
\begin{equation}
\left[\begin{array}{ccccc}
0 & a\gamma & 0 & \cdots & 0\\
P_{1} & -Q_{1} & R_{1} & \cdots & 0\\
0 & P_{2} & -Q_{2} & \cdots & 0\\
\vdots & \vdots & \vdots & \ddots & R_{n-1}\\
0 & 0 & 0 & P_{n} & -Q_{n}
\end{array}\right].
\nonumber
\end{equation}
In this instance,
\begin{equation}
H_{1}=H\left(a,\, q_{n,m};\,-n,\,\beta,\,\gamma,\,\delta,\,\xi\right)\label{eq:Heun_poly}
\end{equation}
is a polynomial of degree $n$, these solutions are the Heun polynomials,
which have attracted a lot of recent attention in relation to various exactly-solvable
quantum mechanics problems 
(see \cite{Hortacsu_arXiv} and references therein for a general review).

Since $w$ and $\kappa$ are positive quantities, in order to satisfy
the termination condition, Eq.~(\ref{eq:cond_1}), $\mathrm{S}_{\mu}$
must take upon the value of $-1$; therefore, Eq.~(\ref{eq:mew}) becomes
\begin{equation}
\mu=\kappa-w.
\nonumber
\end{equation}
The first termination condition, Eq.~(\ref{eq:cond_1}), also requires
\begin{equation}
\kappa=w-\frac{n+1}{2}.
\label{eq:def_kappa}
\end{equation}
Therefore the exponent parameters become
\[
\alpha=-n,\qquad\beta=2w-n-1,\qquad\gamma=\delta=2w-n
\]
and the accessory parameter becomes
\begin{equation}
q_{n,m}=iS_{E}2w\sqrt{\Delta^{2}-\left(w-\frac{n+1}{2}\right)^{2}}-n\left(w-\frac{n+1}{2}\right).
\label{eq:q_nm}
\end{equation}
When $x\rightarrow-\infty$, $\xi\rightarrow0$; therefore, the right hand side of Eq.~(\ref{eq:Heun_poly}) equals unity. The right hand side of Eq.~(\ref{eq:Heun_poly}) can be expressed as $H\left(1-a,-q_{nm}-\beta n;-n,\beta,\delta,\gamma;1-\xi\right)$ \cite{Maier_07}; therefore, as $x\rightarrow\infty$, $\xi\rightarrow1$ and the Heun polynomial tends to unity; therefore, it can be seen from Eq.~(\ref{eq:first_wave}) that the boundary condition $\Psi_{1,2}\left(\pm\infty\right)=0$ requires $\kappa$ to exceed zero, thus, we obtain the condition that $w>\frac{\left(n+1\right)}{2}$. It should be noted that this puts an upper limit on $n$, the order of termination of the Heun polynomial.
From Eqs.~(\ref{eq:kappa},\ref{eq:def_kappa}) the exact Dirac energy
spectrum is found to be
\begin{equation}
E_{n,m}=\pm\sqrt{\Delta_{n,m}^{2}-\left(w-\frac{n+1}{2}\right)^{2}},
\label{eq:Energy_spectrum}
\end{equation}
where $\Delta_{n,m}$ is a function of $w$ and $n$ and is found
via the satisfaction of the second termination condition, Eq.~(\ref{eq:cond_2}).
Let us first consider the case of $E=0$, in this instance the termination
condition, Eq.~(\ref{eq:cond_2}), is satisfied when $q_{n,m}=0$,
which requires $w=\left(n+1\right)/2$, resulting in unbound states
since in this instance $\Delta_{n,m}=0$, or when
\begin{equation}
q_{n,m}=-n\left(w-\frac{n+1}{2}\right)=\frac{\alpha\beta}{2},
\end{equation}
which requires $\Delta=\pm\left(w-\frac{n+1}{2}\right)$ and in
this case the right hand side of Eq.~(\ref{eq:Heun_poly}) becomes
\begin{equation}
H\left(\frac{1}{2},\,\frac{\alpha\beta}{2};\,\alpha,\,\beta,\,\gamma,\,\gamma,\,\xi\right).\label{eq:Heun_with_n}
\end{equation}
Using the identity \cite{Maier_07} $H\left(a,\, q;\,\alpha,\,\beta,\,\gamma,\,\delta;\, z\right)
=H\left(\frac{1}{a},\,\frac{q}{a};\,\alpha,\,\beta,\,\gamma,\,\alpha+\beta+1-\gamma-\delta;\,\frac{z}{a}\right)$,
allows Eq.~(\ref{eq:Heun_with_n}) to be re-expressed as
\begin{equation}
H\left(2,\,\alpha\beta;\,\alpha,\,\beta,\,\gamma,\,\alpha+\beta+1-2\gamma;\,2\xi\right),
\nonumber
\end{equation}
which reduces to the Gauss hypergeometric function \cite{Maier_05}
\begin{equation}
\,_{2}F_{1}\left(\frac{1}{2}\alpha,\frac{1}{2}\beta;\,\gamma;\,4\xi\left(1-\xi\right)\right).
\nonumber
\end{equation}
In order to terminate the hypergeometric series and therefore obtain
bound solutions it is necessary to satisfy the condition $\alpha=-2N$,
where $N$ is a positive integer, therefore,
\begin{equation}
\Delta=\pm\left(w-N+\frac{1}{2}\right),
\nonumber
\end{equation}
where $n=2N$ which restores the results obtained in Ref.~\onlinecite{Hartmann_1}. It should be noted that the condition $w>\frac{\left(n+1\right)}{2}$, puts an upper limit on $n$, the order of termination of the Heun polynomial. Notably the first mode occurs at $n=0$, thus there is a lower threshold of $w>\frac{1}{2}$ for which bound modes appear. Hence within graphene, quantum wells are very different to the non-relativistic case; bound states are not present for any symmetric potential, they are only present for significantly strong or wide potentials, such that $V_{0}l>\frac{1}{2}$.

\begin{figure}[h]
\includegraphics*[height=80mm,angle=270]{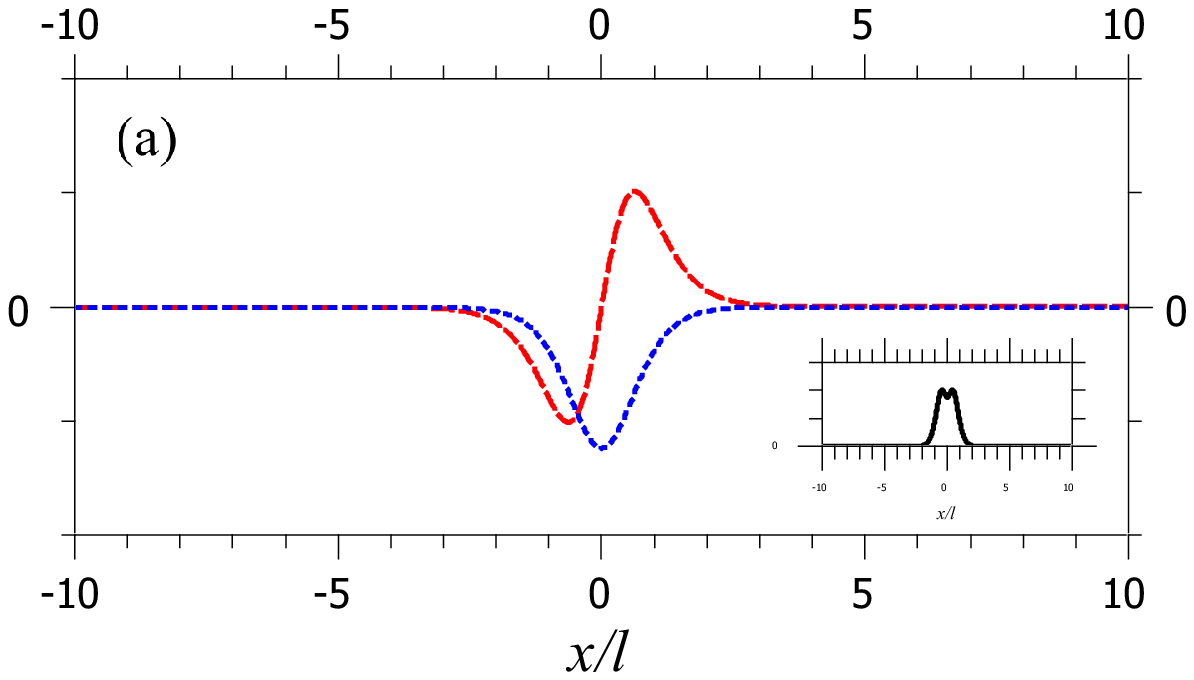}
\includegraphics*[height=80mm,angle=270]{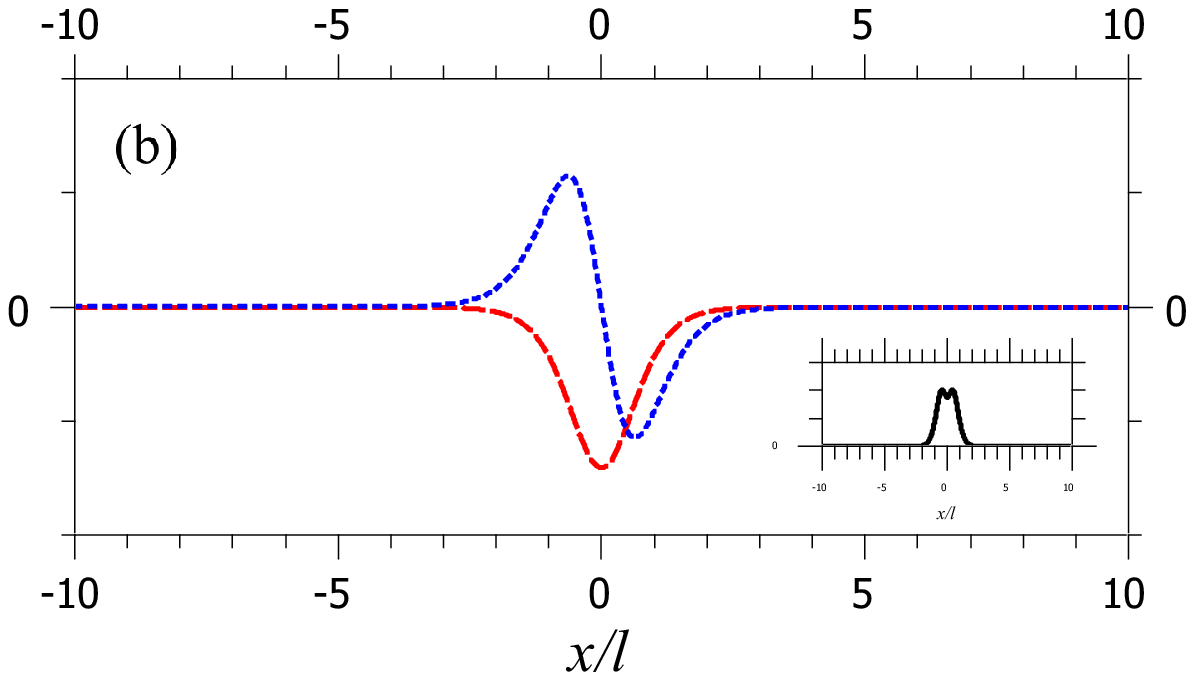}\\
\includegraphics*[height=80mm,angle=270]{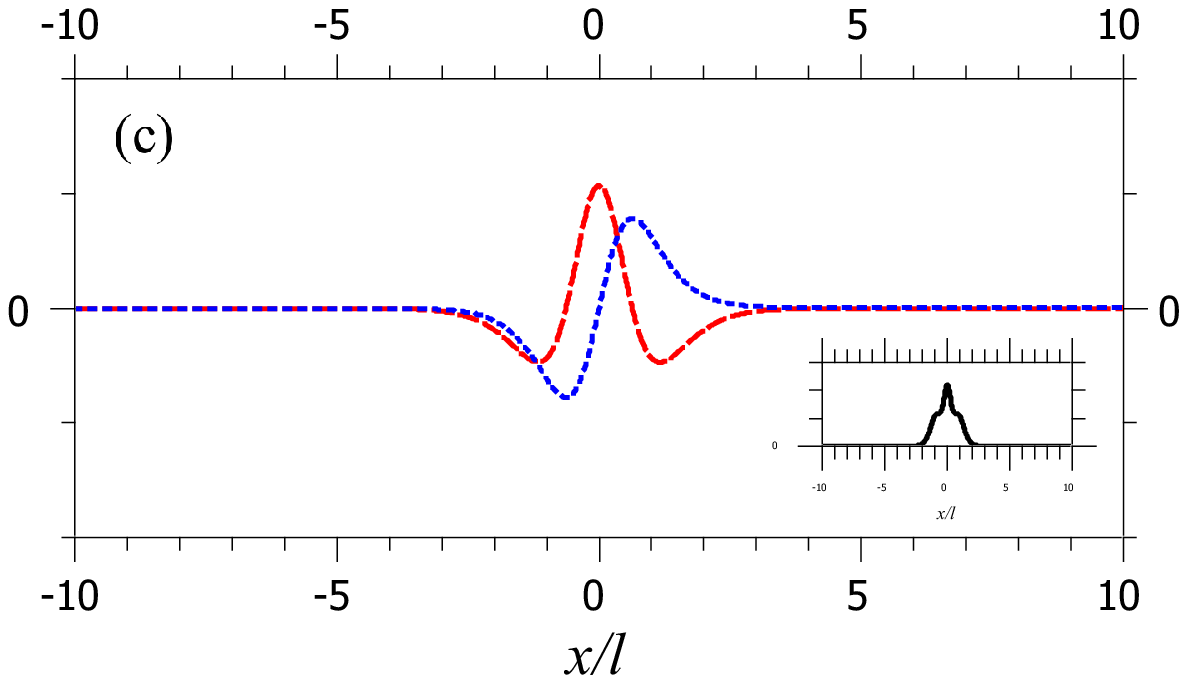}
\includegraphics*[height=80mm,angle=270]{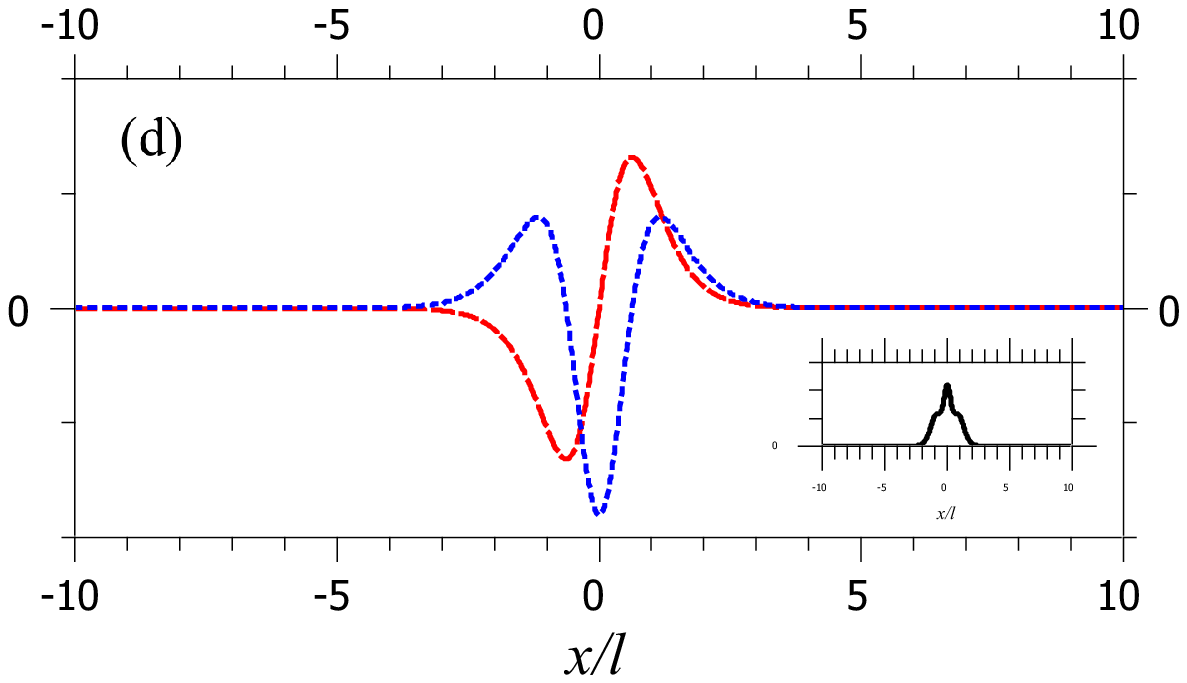}\\
\caption{(Color online)
The real part of the wavefunctions $\Psi_{\mathrm{I}}$ (red long-dashed line) and $\Psi_{\mathrm{II}}$ (blue short-dashed line) are shown for $\omega=3.2$ for: (a) the $E_{1,0}$ mode with $\Delta>0$, (b) the $E_{1,0}$ mode with $\Delta<0$, (c) the $E_{1,1}$ mode with $\Delta>0$ and (d) the $E_{1,1}$ mode with $\Delta<0$. The insets show the electron density profile for the corresponding modes.
}
\label{fig1}
\end{figure}

The non-zero exact energy eigenvalues are obtained by solving Eq.~(\ref{eq:q_nm}).
When $n=0$ the eigenvalue is found to be $E_{0,0}=0$, where $\Delta_{0,0}=\pm(w-\frac{1}{2})$.
For the case of $n=1$, which exists only when $w$, the characteristic potential
strength, exceeds one, the eigenvalues are
\begin{equation}
E_{1,0}=-\frac{1}{2w}\sqrt{w^{2}-w}
\nonumber
\end{equation}
and
\begin{equation}
E_{1,1}=\frac{1}{2w}\sqrt{w^{2}-w},
\nonumber
\end{equation}
where
\begin{equation}
\Delta_{1,0}=\Delta_{1,1}=\pm\frac{2w-1}{2w}\sqrt{w^{2}-w}.
\nonumber
\end{equation}
Bound states of both positive and negative energies exist in the energy
spectrum, which is markedly different to quantum wells in the non-relativistic
case. Each eigenvalue is two-fold degenerate in terms of $\Delta$,
the particles momentum along the barrier.
In Fig.~\ref{fig1} we present $\Psi_{\mathrm{I}}$, $\Psi_{\mathrm{II}}$ and the corresponding electron density profiles for the $E_{10}$ and $E_{11}$ modes.
It can be seen from Fig.~\ref{fig1} that upon changing the sign of $\Delta$ the parity of $\Psi_{\mathrm{I}}$ and $\Psi_{\mathrm{II}}$ changes.
This means backscattering within a channel requires a change in the parity of the wavefunctions and thus should be strongly suppressed.
Such suppression should result in an increase in the mean free path of the channel compared to that of graphene.
Each component of the spinor wavefunction acts much like the single component wavefunction of a conventional quantum well; when $\Delta>0$ ($\Delta<0$), $\Psi_{\mathrm{II}}$ ($\Psi_{\mathrm{I}}$) for the lowest energy state, $E_{10}$, is s-like and for the next excited state, $E_{11}$, p-like. Since $\Psi_{\mathrm{I}}$ ($\Psi_{\mathrm{II}}$) is the derivative of $\Psi_{\mathrm{II}}$ ($\Psi_{\mathrm{I}}$), it must be p-like for $E_{10}$ and d-like for $E_{11}$. The $E_{10}$ mode has a dip in the charge density profile at the middle of the potential well, whereas the $E_{11}$ mode has a maximum. These counterintuitive density profiles arise from the complex two-component structure of the wavefunctions.

\begin{figure}[h]
\includegraphics*[height=100mm,angle=270]{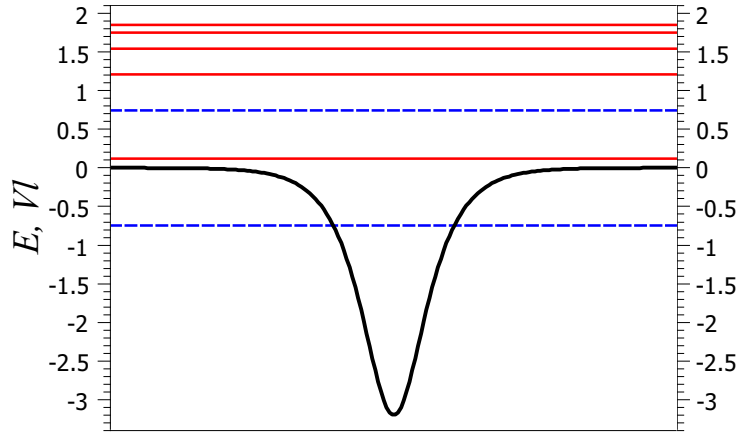}
\caption{(Color online)
Schematic energy spectrum of $-w/\cosh\left(x/l\right)$, for the case of $w=3.2$ and $\Delta=1.856$, in this instance there are 7 eigenvalues. The solid (red) and long-dashed (blue) lines correspond to the calculated and exactly determined eigenvalues respectively. The potential profile is shown in the same scale.
}
\label{fig3}
\end{figure}

The bound modes which propagate along the potential well each contribute $4e^{2}/{h}$ to the channels conductance, where the factor of four accounts for the valley and spin degeneracy. By modulating the parameters of the potential and or changing the position of the Fermi level one can increase the conductance of the channel by multiples of $4e^{2}/{h}$ therefore a change of geometry, from normal transmission to propagation along a potential, allows graphene to be used as a switching device. The existence of bound modes within smooth potentials in graphene may provide an additional argument in favor of the mechanism for minimal conductivity, where charge puddles lead to a percolation network of conducting channels \cite{PercolationNetwork}.
It can be see from Eq.~(\ref{eq:Energy_spectrum}) 
that the exact solutions correspond to the case where there exists a bound state at equal energy above and below the top of the well. For example, a potential of characteristic strength $w=3.2$ with $\Delta=1.856$ contains 7 bound modes, two of which are symmetric about $E=0$, as shown in Fig.~\ref{fig3}.

\section{Conclusions}
We have presented the hitherto unknown quasi-exact solutions to the Dirac equation for the hyperbolic secant potential, which provides a good fit for potential profiles of existing top-gated graphene structures. It was found that bound states of both positive and negative energies exist in the energy spectrum and that there is a threshold value of $w$, the characteristic potential strength, for which the first mode appears. 

\section*{Acknowledgements}
We are grateful to Charles Downing for valuable discussions and thank Katrina Vargas and Elvis Arguelles for the critical reading of the manuscript. This work was supported by URCO (17 N 1TAY12-1TAY13), the EU FP7 ITN NOTEDEV (Grant No. FP7-607521) and FP7 IRSES projects SPINMET (Grant No. FP7-246784), QOCaN (Grant No. FP7-316432), and InterNoM (Grant No. FP7-612624).

\section*{Appendix}
List of eigenvalues and their corresponding $\Delta_{n,m}$

~

$E_{0,0}=0$

$E_{1,0}=-\frac{1}{2w}\sqrt{w^{2}-w}$

$E_{1,1}=-E_{1,0}$

$E_{2,0}=-\frac{1}{2w}\sqrt{4w^{2}-6w+1}$

$E_{2,1}=0$

$E_{2,2}=-E_{2,0}$

$E_{3,0}=-\frac{1}{2w}\sqrt{5w^{2}-10w+3+\sqrt{16w^{4}-64w^{3}+85w^{2}-42w+9}}$

$E_{3,1}=-\frac{1}{2w}\sqrt{5w^{2}-10w+3-\sqrt{16w^{4}-64w^{3}+85w^{2}-42w+9}}$

$E_{3,2}=-E_{3,1}$

$E_{3,3}=-E_{3,0}$

$E_{4,0}=-\frac{1}{4w}\sqrt{40w^{2}-100w+42+6\sqrt{16w^{4}-80w^{3}+140w^{2}-100w+33}}$

$E_{4,1}=-\frac{1}{4w}\sqrt{40w^{2}-100w+42-6\sqrt{16w^{4}-80w^{3}+140w^{2}-100w+33}}$

$E_{4,2}=0$

$E_{4,3}=-E_{4,1}$

$E_{4,4}=-E_{4,0}$

~

$\Delta_{0,0}=\pm\left(w-\frac{1}{2}\right)$

$\Delta_{1,0}=\Delta_{1,1}=\pm\frac{1}{w}\sqrt{w^{2}-w}\left(w-\frac{1}{2}\right)$

$\Delta_{2,1}=\pm\left(w-\frac{3}{2}\right)$

$\Delta_{2,0}=\Delta_{2,2}=\pm\frac{1}{2w}\left(2w^{2}-3w+1\right)$

$\Delta_{3,0}=\Delta_{3,3}=\pm\frac{1}{2w}\sqrt{4w^{4}-16w^{3}+21w^{2}-10w+3+\sqrt{16w^{4}-64w^{3}+85w^{2}-42w+9}}$

$\Delta_{3,1}=\Delta_{3,2}\pm\frac{1}{2w}\sqrt{4w^{4}-16w^{3}+21w^{2}-10w+3-\sqrt{16w^{4}-64w^{3}+85w^{2}-42w+9}}$

$\Delta_{4,2}=\pm\left(w-\frac{5}{2}\right)$

$\Delta_{4,0}=\Delta_{4,4}=\pm\frac{1}{2w}\left(\frac{3}{2}+\frac{1}{2}\sqrt{16w^{4}-80w^{3}+140w^{2}-100w+33}\right)$

$\Delta_{4,1}=\Delta_{4,3}=\pm\frac{1}{2w}\left(\frac{3}{2}-\frac{1}{2}\sqrt{16w^{4}-80w^{3}+140w^{2}-100w+33}\right)$

\end{document}